\documentclass[11pt,twocolumn,english,american,aip,jap,numerical,reprint,groupedaddress]{revtex4-1}

\usepackage[T1]{fontenc}
\usepackage[latin9]{inputenc}
\usepackage{color}
\usepackage{babel}
\usepackage{float}
\usepackage{bm}
\usepackage{amsmath}
\usepackage{amsthm}
\usepackage{amssymb}
\usepackage{graphicx}
\usepackage{wasysym}
\usepackage{esint}
\usepackage[unicode=true,pdfusetitle,
 bookmarks=true,bookmarksnumbered=false,bookmarksopen=false,
 breaklinks=true,pdfborder={0 0 1},backref=false,colorlinks=true]
 {hyperref}
\usepackage{breakurl}
\usepackage{chngcntr}

\makeatletter

\@ifundefined{textcolor}{}
{%
 \definecolor{BLACK}{gray}{0}
 \definecolor{WHITE}{gray}{1}
 \definecolor{RED}{rgb}{1,0,0}
 \definecolor{GREEN}{rgb}{0,1,0}
 \definecolor{BLUE}{rgb}{0,0,1}
 \definecolor{CYAN}{cmyk}{1,0,0,0}
 \definecolor{MAGENTA}{cmyk}{0,1,0,0}
 \definecolor{YELLOW}{cmyk}{0,0,1,0}
}

\usepackage{color}
\usepackage{babel}
\usepackage{bm}
\usepackage{breakurl}

\providecommand{\second}{\prime\prime}

\@ifundefined{textcolor}{}{%
 \definecolor{BLACK}{gray}{0}
 \definecolor{WHITE}{gray}{1}
 \definecolor{RED}{rgb}{1,0,0}
 \definecolor{GREEN}{rgb}{0,1,0}
 \definecolor{BLUE}{rgb}{0,0,1}
 \definecolor{CYAN}{cmyk}{1,0,0,0}
 \definecolor{MAGENTA}{cmyk}{0,1,0,0}
 \definecolor{YELLOW}{cmyk}{0,0,1,0}
}

\AtBeginDocument{
  
}

\makeatother

\begin{document}

\title{Specific absorption rate of magnetic nanoparticles: nonlinear AC
susceptibility}

\author{J.-L. Déjardin}
\email{jean-louis.dejardin@orange.fr}

\author{F. Vernay}
\email{francois.vernay@univ-perp.fr}

\author{H. Kachkachi}
\email{hamid.kachkachi@univ-perp.fr}

\date{\today}
\begin{abstract}
In the context of magnetic hyperthermia, several physical parameters
are used to optimize the heat generation and these include the nanoparticles
concentration and the magnitude and frequency of the external AC magnetic
field. 
Here we extend our previous work by computing nonlinear contributions to the
specific absorption rate, while taking into account (weak) inter-particle
dipolar interactions and DC magnetic field. In the previous work, the latter were shown to
enhance the SAR in some specific geometries and setup. We find that the cubic
correction to the AC susceptibility does not modify the qualitative
behavior observed earlier but does bring a non negligible quantitative
change of specific absorption rate, especially at relatively high
AC field intensities. 
Incidentally, within our approach based on the AC susceptibility, we revisit the physiological empirical criterion on the upper limit of the product of the AC magnetic field intensity $H_{0}$ and its frequency $f$, and provide a physicist's rationale for it. 
\end{abstract}

\address{Laboratoire PROMES-CNRS (UPR-8521) \& Université de Perpignan Via
Domitia, Rambla de la thermodynamique, Tecnosud, 66100 Perpignan,
FRANCE}
\maketitle

\section{Introduction}

Magnetic hyperthermia is a promising route for cancer therapy which
consists in injecting a low treatment dose of magnetic nanoparticles
(NP) in the targeted cells. The NP are heated with the help of an
AC magnetic field and the optimization of the whole process depends
on several factors, such as the magnetic field itself (in strength
and frequency), NP size and concentration, and solution viscosity.\citep{Carrey_JAP2011,Mehdaoui_AFM2011,martinez2013learning,condeetal15jpcc,koslap2015nanotechrev}
From the fundamental point of view, a measure of the efficiency of
magnetic hyperthermia is provided by the so-called \emph{specific
absorption rate} (SAR), usually given in Watt per gram (W/g). In order
to avoid possible side effects, the smallest possible dose of NP has
to be introduced in the human body and, as such, it is very important
to assess the effect of NP concentration and the ensuing inter-particle
interactions. Then, it has been shown by many authors that magnetic
heating is enhanced by increasing the NP size. However, there are
also limitations on the maximum size of NPs that can be administered
to humans.\citealp{SHAH201596} The AC magnetic field (of intensity
$H_{0}$ and frequency $f$) offers then another handle on the optimization
of heat generation and enhancement of the SAR. Here again a limit
on the patient tolerance, for an exposed area of a given size, has been defined as a limit on the product\citealp{brezovich88mpm}
$H_{0}f\simeq4.85\times10^{8}\,{\rm A.m^{-1}.s^{-1}}$. Several works
have demonstrated that heating rates and SAR values increase with
the intensity of the magnetic field.\citealp{gloverEtal_ieee13,bordelonEtal_jap11,muraseEtal_jap11,gonzalezEtal_j3m09,guardiaEtal_acnano12}
Thus, to allow for a wider range of variation of the AC field intensity,
at low frequency, one has to investigate the dynamic response of the
NP assembly for large values of $H_{0}$.

The effect of all these physical parameters has been studied by many
research groups by computing the hysteresis loop of the magnetization
as a function of the AC field intensity, for a given frequency.\citealp{condeetal15jpcc,martinez2013learning,Carrey_JAP2011,Lacroix_etal_JAP2009,Mehdaoui_AFM2011,mehdaouietal12apl}
The SAR is then inferred from these calculations as being proportional
to the area of the hysteresis loop. The latter is a physical observable
that emerges owing to a lag of the system's response, the magnetization,
with respect to the excitation, here the AC field. This is due
to the fact that in the process the system goes through several metastable
states. It is then not easy, if not impossible, to perform an analytical
study of the hysteretic magnetization as a function of the applied
field, in the general situation. As a consequence, it is not possible to derive analytical expressions 
for the SAR when inferred from the hystereis loop. 
On the other hand, analytical developments are possible through the alternative
approach based on the fact that the system absorption of the electromagnetic
energy brought in by the AC field is described by the out-of-phase
component of the dynamic response function, the AC susceptibility.
Indeed, it is well known that the absorption (dissipation) of energy
can, in general, be described by the imaginary part of the permittivity
and permeability of the medium. More precisely, for a monochromatic
magnetic field, the time average of $-\bm{\nabla}\cdot\bm{S}$, where
$\bm{S}$ is the Poynting vector, yields the average heat $Q$ dissipated
in the medium per unit time and unit volume. $Q$ can be written as\citep{lanlif8}
\begin{equation}
Q=\frac{\omega}{4\pi}\left[\epsilon^{\prime\prime}\overline{\bm{E}^{2}}+\mu^{\prime\prime}\overline{\bm{H}^{2}}\right]\label{eq:DissipatedHeat}
\end{equation}
where $\omega$ is the angular frequency of the AC field. $\bm{E}$ and $\bm{H}$ are the real amplitudes of the electric
and magnetic field and the bar stands for time average. $\epsilon^{\prime\prime}$
and $\mu^{\prime\prime}$ are respectively the imaginary parts of
the permittivity and permeability of the medium. Eq. (\ref{eq:DissipatedHeat})
shows that the heat dissipated in the system is proportional to the
field frequency and to the square of its amplitude. This result does
not exclude situations where $\epsilon^{\prime\prime}$ and $\mu^{\prime\prime}$
are functions of the applied fields\citep{lanlif8}. Indeed, focusing on the magnetic
field contribution and expanding the magnetic permeability in powers
of the magnetic field $H$, \emph{i.e.,} $\mu=1+\chi=1+\chi^{\left(1\right)}+\chi^{\left(3\right)}H^{2}+\ldots\equiv\mu_{1}+\chi^{\left(3\right)}H^{2}+\ldots$,
one can show that $Q\simeq\frac{\omega}{4\pi}\overline{\left[\mu_{1}^{\prime\prime}+\left(\chi^{\left(3\right)}\right)^{\prime\prime}H^{2}\right]H^{2}}$,
with the first correction to $Q$ of order $4$ in $H$ with a coefficient
given by the imaginary part of the cubic susceptibility.

In the context of magnetic hyperthermia, averaging over one cycle
of the AC magnetic field, $\bm{H}_{AC}=H_{0}\exp\left(i\omega t\right)\bm{e_{x}}$,
yields the energy dissipated per cycle. The SAR is then shown to be
directly proportional to the imaginary component of the AC susceptibility
$\chi^{\prime\prime}\left(\omega\right)$.\citep{rosensweig02j3m,Hergt_etal_JPMC2006,Ahrentorp_etal_aipcp2010}
More precisely, we have
\begin{equation}
{\rm SAR}=\frac{\mu_{0}\omega}{2}H_{0}^{2}\chi^{\prime\prime}\left(\omega\right).\label{eq:SAR_chi_ac}
\end{equation}

In a previous work\citep{dejardin_etal_SAR_JAP_2017}, we used the
expression above and studied the effect on SAR of the inter-particle
interactions and DC magnetic field with the help of available analytical
expressions for the imaginary part of the AC susceptibility,\citep{garpal00acp,Vernay_etal_acsucept_PRB2014}
in the linear approximation. The value used for $H_{0}$ was about
$7.3\,{\rm mT}$ with a frequency $f=56\ {\rm kHz}$. Now, several
experimental studies are carried out for much larger values of the
AC field intensity. For instance, in Ref. \onlinecite{SHAH201596}
the effect of the AC field is investigated for an intensity in the
range: $15.1-47.7\,{\rm kA/m}$ or equivalently $19-60\,{\rm mT}$.
This then addresses the question as to whether the linear approximation,
with respect to the AC field amplitude, can still be used. To answer this 
question one has to compute the contribution to the SAR from the nonlinear
terms in the AC susceptibility $\chi^{\prime\prime}\left(\omega\right)$,
especially the next order, \emph{i.e.} the cubic susceptibility, as
discussed above. Incidentally, this could help improve the quantitative
comparison with experiments, in addition to the already good qualitative
agreement reported in Ref. \onlinecite{dejardin_etal_SAR_JAP_2017}.
This is the main objective of the present work.

Nonlinear AC susceptibility for magnetic nanoparticles has been studied
by many authors.\citep{BitohEtal_JPSJ.64.1311,BitohEtal_J3M.154.59,RaikherStepanov_PhysRevB.55.15005,RaikherStepanov_JMMM.196.88,JonssonEtAl_JMMM.222.219,RaikherStepanov_PhysRevB.66.214406,garpalgar04prb,WangHuang_CPL.421.544,FickoEtAl_JMMM.378.267}
In particular, the work by Raikher and Stepanov\citep{RAIKHER_JMMM2008} highlights the importance of nonlinear effects at high-field amplitudes, even in the context of Brownian rotation in a ferrofluid. In fact, as discussed by Vallejo-Fernandez \emph{et al.}\citep{OGRADY_JPHYSD2013,OGRADY_APL2013}, three heating mechanisms may contribute to the SAR: susceptibility loss, hysteresis loss and viscous heating, the first one dominates for small sizes. The authors argue that due to the variety of physical parameters and mechanisms that influence the SAR in {\it in vivo} samples, it is instructive to investigate solid matrices since this allows us to simplify the study by getting rid of the mechanical rotation of the nanoparticles and viscous heating. Now, we stress that already in solid samples, it is rather difficult to build analytical models in the general situation of arbitrary anisotropy, applied field, temperature and damping parameter.

In the remainder of the paper, we will apply the formalism of
Ref. [\onlinecite{garpalgar04prb}] to investigate
the effects of nonlinear AC susceptibility and their competition with
the dipolar interactions in the specific absorption rate of an assembly
of monodisperse magnetic nanoparticles with oriented uniaxial anisotropy,
in a longitudinal DC magnetic field. Our investigation thus takes into account
the effect of the competition between the DC field and the
dipolar field. In particular, we will compare
the linear and nonlinear contributions to the SAR as we increase the
AC field amplitude.

The paper is organized as follows: in the next Section we present
our model and hypotheses and in Section \ref{sec:Susceptibility},
the first nonlinear correction to the susceptibility and to the SAR
is derived. The results are discussed in Section \ref{sec:Results},
which is followed by our conclusions.

\section{\label{sec:Model-and-Hypotheses}Model and Hypotheses}

\begin{figure}[H]
\begin{centering}
\includegraphics[width=6.5cm]{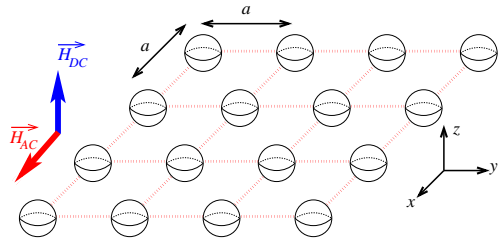}
\par\end{centering}
\centering{}\caption{\label{fig:Oblate-2D-sample}2D square assembly of nano-spheres on
a square super-lattice of parameter $a$ (same geometry as in Ref. [\onlinecite{dejardin_etal_SAR_JAP_2017}]). The external DC field is
applied along the $z$-axis, the AC field lies in the $xy$-plane. The assembly is monodisperse and textured with all anisotropy easy
axes oriented in the $z$ direction.}
\end{figure}

As stated earlier, several physical parameters influence the SAR (nanoparticles
size, material, temperature, field intensity, frequency,...). The
aim of the present work is not to provide a systematic investigation
of the whole parameter space, but rather to pinpoint the role of the
nonlinearities of the magnetic susceptibility. Hence, in order to
make a consistent quantitative analysis, we consider, as in Ref. \onlinecite{dejardin_etal_SAR_JAP_2017},
a monodisperse assembly of $\mathcal{N}$ single-domain nanoparticles
with oriented uniaxial anisotropy, each carrying a magnetic moment
$\bm{m}_{i}=m_{i}{\bf s}_{i},\,i=1,\cdots,{\cal N}$ of magnitude
$m$ and direction ${\bf s}_{i}$, with $\vert{\bf s}_{i}\vert=1$.
Each nanoparticle of volume $V$ is attributed an (effective) uniaxial
anisotropy constant $K_{{\rm eff}}$ with an easy-axis in the $z$
direction. The magnetic moments $\bm{m}_{i}$ are located at the vertices
of a simple $2D$ square super-lattice of parameter $a$, in the $xy$
plane. The geometry of the system is sketched in Fig. \ref{fig:Oblate-2D-sample}.

In the present work, we consider the same system
setup as in Ref. \onlinecite{dejardin_etal_SAR_JAP_2017}, namely
of oriented uniaxial anisotropy for all nanoparticles and a longitudinal
DC magnetic field, in the intermediate--to-high damping regime, and
in the high-anisotropy limit. In this case, the cubic contribution
to the AC susceptibility is given by Eq. (37) of Ref. \onlinecite{garpalgar04prb},
\emph{i.e.}

\begin{equation}
\chi^{\left(3\right)}=\chi_{{\rm eq}}^{\left(3\right)}\frac{1-\frac{1}{2}i\eta}{\left(1+i\eta\right)\left(1+3i\eta\right)}\label{eq:CubicSuscep}
\end{equation}

where 
\[
\chi_{{\rm eq},{\rm free}}^{\left(3\right)}=-\frac{1}{3}\frac{m^{4}}{\left(k_{B}T\right)^{3}}.
\]
and $\eta=\omega\tau$, $\tau$ is the (longitudinal) N\'eel relaxation time, and $\chi_{{\rm eq}}^{\left(3\right)}$ is the cubic contribution
to the equilibrium susceptibility.
The high-energy barrier limit requires $\sigma=K_{{\rm eff}}V/k_{B}T\gg1$, where
$K_{{\rm eff}}$ is considered to be the largest energy scale in the
present calculations and $k_B$ is the Boltzmann constant. This limit applies in the context of hyperthermia
experiments (at temperature $T\simeq318{\rm K}$). Indeed, for iron-cobalt
nanoparticles of volume $V\sim5.23\times10^{-25}{\rm m^{3}}$ (\emph{i.e.}
spheres of radius $R=5\ {\rm nm}$), with an effective anisotropy
constant $K_{{\rm eff}}\sim4.5\times10^{4}{\rm J.m^{-3}}$ and a density
$\rho\simeq8300\ {\rm kg.m^{-3}}$, we have $\sigma\simeq5.4$.

Taking account of dipolar interactions (DI) together with uniaxial
anisotropy and DC magnetic field ($\bm{H}_{{\rm ex}}=H_{{\rm DC}}\bm{e_{z}}$),
the energy of a magnetic moment $\bm{m}_{i}$ within the assembly, reads (after multiplying by $-\beta\equiv-1/k_{B}T$)\citep{dejardin_etal_SAR_JAP_2017}

\begin{equation}
\mathcal{E}_{i}=x\,{\bf s}_{i}\cdot\bm{e_{z}}+\sigma\left(\mathbf{s}_{i}\cdot\bm{e_{z}}\right)^{2}+{\cal E}_{i}^{\mathrm{DI}},\label{eq:DDIAssemblyEnergy}
\end{equation}
where $x=\beta mH_{{\rm DC}}$ and ${\cal E}_{i}^{\mathrm{DI}}$ is
the contribution from the long-range DI, 
\begin{equation}
{\cal E}_{i}^{\mathrm{DI}}=\xi\sum_{j<i}{\bf s}_{i}\cdot{\cal D}_{ij}\cdot{\bf s}_{j},\label{eq:Energy-DIcontribution}
\end{equation}
with the usual notation: ${\cal D}_{ij}=\left(3\bm{e}_{ij}\bm{e}_{ij}-1\right)/r_{ij}^{3}$,
$\bm{r}_{ij}=\bm{r}_{i}-\bm{r}_{j}$, $r_{ij}=\left|\bm{r}_{ij}\right|,\bm{e}_{ij}=\bm{r}_{ij}/r_{ij}$.
For convenience, we also use the dimensionless DI coefficient $\xi$
\begin{equation}
\xi=\left(\frac{\mu_{0}}{4\pi}\right)\left(\frac{m^{2}/a^{3}}{k_{B}T}\right).\label{eq:DI-xi}
\end{equation}

In summary, the variable energy parameters of our calculations are $\xi$ and
$x$, in addition to the AC field amplitude $H_{0}$ ; the frequency
$\omega=2\pi f$ of the latter and the remaining parameters will be
held constant. We will adopt a perturbative approach to derive analytical
expressions for the SAR as a function of $\xi,x,H_{0}$ and use them
to investigate the effect of $x,\xi$, as we vary $H_{0}$ in the
wide range explored by experiments.

\section{AC Susceptibility \label{sec:Susceptibility} }

The AC susceptibility is a complex quantity that can be written as
$\chi\left(x,\sigma,\xi,\omega\right)=\chi^{\prime}-i\chi^{\prime\prime}$.
As usual, it can be expanded in terms of the amplitude $H_{0}$ of
the AC field\citep{RaikherStepanov_PhysRevB.55.15005}

\begin{eqnarray}
\chi & = & \chi^{\left(1\right)}+3H_{0}^{2}\chi^{\left(3\right)}+\ldots\ .\label{eq:XitotalAssembly}
\end{eqnarray}
$\chi^{\left(1\right)}$ is the linear contribution which, according
to the model of Debye, reads

\begin{equation}
\chi^{\left(1\right)}=\chi_{\mathrm{eq}}^{\left(1\right)}\frac{1}{1+i\eta},\label{eq:Debye}
\end{equation}
where $\chi_{\mathrm{eq}}^{\left(1\right)}$ is the equilibrium susceptibility.
According to the Debye theory\citep{debye1929Dover,zwanzig_jcp63p2766}
the only interaction of the molecule (here a nanoparticle) is with
the external field. In the context of magnetic nanoparticles, this
theory describes the absorption by a single mode of the electromagnetic
energy brought in by the external field. In the case of weak inter-particle
interactions, the dynamics of this mode is rather slow and characterized
by the longitudinal relaxation time $\tau$, corresponding to the
population inversion from the blocked state to the superparamagnetic
state. The latter transition corresponds on average to the crossing
by each nanoparticle\textquoteright s magnetic moment of its energy
barrier. As argued in Ref. \onlinecite{zwanzig_jcp63p2766}, dipolar
interactions introduce new relaxation times in higher-order perturbation
theory and are associated with further losses at higher frequencies.
However, in the hyperthermia context, the magnetic field frequency
$\omega$ is relatively low (a few hundred kHz) and as such the Debye
approximation with the first term in Eq. (\ref{eq:Debye}) is sufficient.
On the other hand, the corrections due to (weak) dipolar interactions
can be taken into account through the equilibrium susceptibility $\chi_{\mathrm{eq}}^{\left(1\right)}$,
see Eqs. (11), (14) and (15) in Ref. \onlinecite{dejardin_etal_SAR_JAP_2017},
within the limit of small DC field and linear equilibrium susceptibility.
More precisely, we introduced $\chi_{{\rm eq,free}}$ the ``free'' contribution
for a single particle and the interacting contribution
$\chi_{{\rm eq,int}}$ to the equilibrium suceptibility, and therefore 
wrote 
\begin{equation}
\chi_{\mathrm{eq}}^{\left(1\right)}=\chi_{{\rm eq,free}}^{\left(1\right)}+\tilde{\xi}\chi_{{\rm eq,int}}^{\left(1\right)},\label{eq:Chi1-eq-Expanded}
\end{equation}
where $\tilde{\xi}=\xi\mathcal{C}^{\left(0,0\right)}$, is the genuine
coefficient that accounts for the DI intensity $\xi$ and the super-lattice
through the lattice sum $\mathcal{C}^{\left(0,0\right)}$ which evaluates
to $\mathcal{C}^{\left(0,0\right)}\simeq-9$ for the square sample
shown in Fig. \ref{fig:Oblate-2D-sample}. 
All the details of the
analytical expressions of these various quantities are available in
Ref. \onlinecite{dejardin_etal_SAR_JAP_2017} and will not be reproduced
here.

In the linear regime with respect to the AC magnetic field, \emph{i.e.
}restricting the expansion in Eq. (\ref{eq:XitotalAssembly}) to the
first term, the SAR was computed in Ref. \onlinecite{dejardin_etal_SAR_JAP_2017}
as a function of the DC magnetic field and assembly concentration,
see Fig. 6 therein. Now, we extend the expansion one step further
and include the second term with the cubic susceptibility $\chi^{\left(3\right)}$
given by (\ref{eq:CubicSuscep}). Consequently, the general expression
of the SAR in Eq. (\ref{eq:SAR_chi_ac}) can then be regarded as a
double expansion: i) an expansion to second order in terms of the
magnitude $H_{0}$ of the AC magnetic field (thus bringing the linear
and cubic susceptibilities, \emph{i.e.} $\chi^{\left(1\right)}$ and
$\chi^{\left(3\right)}$), and ii) an expansion in the DI strength
$\tilde{\xi}$ for the equilibrium susceptibilities ($\chi_{\mathrm{eq}}^{\left(1\right)}$
and $\chi_{\mathrm{eq}}^{\left(3\right)}$). So, in principle one
should perform the latter expansion for both $\chi_{\mathrm{eq}}^{\left(1\right)}$
and $\chi_{\mathrm{eq}}^{\left(3\right)}$. However, writing $\chi_{{\rm eq}}^{\left(3\right)}=\chi_{{\rm eq,free}}^{\left(3\right)}+\tilde{\xi}\chi_{{\rm eq,int}}^{\left(3\right)}$,
similarly to Eq. (\ref{eq:Chi1-eq-Expanded}), leads to ($\Im$ stands
for imaginary part)

\begin{align}
{\rm SAR} & \simeq\frac{\mu_{0}\omega}{2\pi}H_{0}^{2}\Im\left[\chi_{{\rm free}}^{\left(1\right)}+\tilde{\xi}\chi_{{\rm int}}^{\left(1\right)}+3H_{0}^{2}\left(\chi_{{\rm free}}^{\left(3\right)}+\tilde{\xi}\chi_{{\rm int}}^{\left(3\right)}\right)\right]\nonumber \\
 & ={\rm SAR}^{\left(1\right)}+\frac{3\mu_{0}\omega}{2\pi}H_{0}^{4}\Im\left(\chi_{{\rm free}}^{\left(3\right)}+\tilde{\xi}\chi_{{\rm int}}^{\left(3\right)}\right)\label{eq:total_SAR}
\end{align}
where ${\rm SAR}^{\left(1\right)}$ is the SAR involving only the
linear susceptibility studied in Ref. \onlinecite{dejardin_etal_SAR_JAP_2017}.

However, we note that at low fields, as it is clearly seen in both the experimental\citep{BitohEtal_JPSJ.64.1311,BitohEtal_J3M.154.59}
and theoretical \citep{RaikherStepanov_PhysRevB.55.15005} results, the imaginary part of $H_{0}^{2}\chi_{{\rm free}}^{\left(3\right)}$
is at least one order of magnitude smaller than that of $\chi_{{\rm eq,free}}^{\left(1\right)}$ (in absolute value), and $\tilde{\xi}\chi_{{\rm int}}^{\left(3\right)}$ is even smaller. 
The calculation of the latter is rather involved and would require, for instance, the generalization of the work by
Raihker and Stepanov\citep{RaikherStepanov_PhysRevB.55.15005} for noninteracting particles in order to include dipolar interactions,
or the extension of the work by Jönsson and Garc\'ia-Palacios to dynamic susceptibility. However, the aim of the present work is not to investigate
concentrated samples, but to determine the general tendency implied by the effects of assemblies on SAR. For these reasons, we
focus on low-values of $\xi$ and neglect higher-order corrections
in $\tilde{\xi}\chi_{{\rm int}}^{\left(3\right)}$. In fact, we think
that it is worth investigating the effect of the cubic AC susceptibility
already through its ``free'' contribution $\chi_{{\rm free}}^{\left(3\right)}$,
remembering that, at any rate, the whole approach assumes weak DI. In conclusion, we will compute the following correction to the SAR
\begin{equation}
{\rm SAR}^{\left(3\right)}\simeq\frac{3\mu_{0}\omega}{2\pi}H_{0}^{4}\Im\left[\chi_{{\rm free}}^{\left(3\right)}\right]\label{eq:CubicSAR}
\end{equation}
where $\chi_{{\rm free}}^{\left(3\right)}$ is given by Eq. (\ref{eq:CubicSuscep}).

More explicitly, we obtain
\begin{align}
{\rm SAR}^{\left(3\right)} & =\frac{3\mu_{0}m^{4}}{4\pi\left(k_{B}T\right)^{3}}H_{0}^{4}\frac{\left(3-\eta^{2}\right)\eta^{2}/\tau}{\left(\eta^{2}+1\right)\left(9\eta^{2}+1\right)}.\label{eq:SAR-nl-Corr}
\end{align}

\section{Results and discussion\label{sec:Results}}

The physical parameters used in the following are basically the same
as those used in our earlier work\citep{dejardin_etal_SAR_JAP_2017}
and have been recalled in Section \ref{sec:Model-and-Hypotheses}.
The DC field is oriented along the anisotropy axis, \emph{i.e.} perpendicular
to the plane, with a normalized magnitude $h=x/2\sigma$.

We decided to focus our investigation on solid samples, with nanoparticles
on the vertices of a square lattice. Other sample geometries could be
investigated by computing the appropriate lattice sum
$\mathcal{C}^{\left(0,0\right)}$, even non-periodic systems can be tackled by
randomly depleting the initial sample. However, in such samples, the main
effect would then be to effectively modulate the dipolar field by changing
the value of $\tilde{\xi}$. The impact of the nonlinearities in these systems
would remain the same since the correction in $\tilde{\xi}\chi_{{\rm int}}^{\left(3\right)}$ are of higher-order in comparison
to the free contribution. 
On the other hand, as discussed earlier, the study of samples with solid matrices makes it possible to simplify the problem by avoiding extra degrees freedom and related loss sources (Brownian motion and field-induced stirring). This is necessary for a more precise and conclusive study of heating due to susceptibility losses, especially for small nanoparticles for which the latter dominate\citep{OGRADY_JPHYSD2013}. In addition, this simplification helps investigate, in a somewhat pure form, nonlinear effects of magnetization processes induced by an increasing AC field.

\begin{figure}
\begin{centering}
\includegraphics[width=6.5cm]{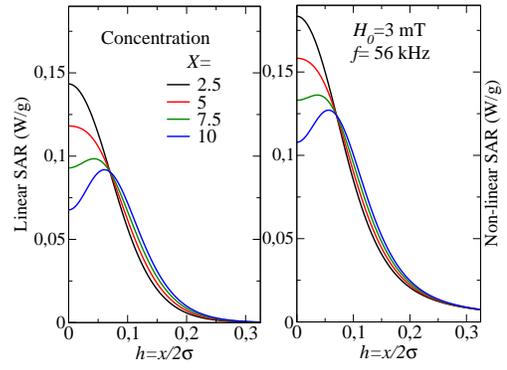}
\par\end{centering}
\caption{Left panel: linear SAR (${\rm SAR^{\left(1\right)}}$) as a function
of the DC field for a 2D sample, for different concentrations $X$.
Right panel: nonlinear SAR (${\rm SAR}={\rm SAR^{\left(1\right)}}+{\rm SAR^{\left(3\right)}}$).
For both panels the AC field intensity is $H_{0}=3\ {\rm mT}$ and
frequency $f=56\ {\rm kHz}$.\label{fig:SAR_vs_dcField-hac=00003D3mT}}
\end{figure}

The first term of expression (\ref{eq:total_SAR}),\emph{ i.e.} ${\rm SAR^{\left(1\right)}}$,
the linear susceptibility contribution, is plotted on the left panel
of Fig. \ref{fig:SAR_vs_dcField-hac=00003D3mT} against the DC field
for different concentrations $X=10^{-21}/a^{3}$ , where the lattice
parameter $a$ is expressed in meters, so an inter-particle separation
of 46 nm corresponds to $X\sim10$. The right panel of Fig. \ref{fig:SAR_vs_dcField-hac=00003D3mT}
displays the behavior of the SAR as expressed in Eq. (\ref{eq:total_SAR})
which includes the first nonlinear correction to the susceptibility
given in Eq. (\ref{eq:SAR-nl-Corr}). From a quantitative point of
view, one can see that the SAR is slightly enhanced by the nonlinear
correction. This is due to the fact that while ${\rm SAR^{\left(1\right)}}$
scales like $H_{0}^{2}$, ${\rm SAR^{\left(3\right)}}$ has a quartic
behavior in terms of the AC field intensity. In the present case,
with such a relatively low AC field intensity ($H_{0}=3\ {\rm mT}$),
neither the quantitative nor the qualitative behavior is notably modified:
at low DC field, for such a square sample, the SAR is reduced by the
DI which compete with the anisoptropy and thus soften the whole magnetic
system. This competition also leads to a non-monotonic behavior of
the SAR as a function of $h$ and a crossing of the curves, as discussed
in a previous work.\citep{dejardin_etal_SAR_JAP_2017} This
result shows that the nonlinear correction to the SAR does not alter
our conclusion that in some specific situations, the SAR may be optimized
by applying an external DC field.

\begin{figure}
\begin{centering}
\includegraphics[width=6.5cm]{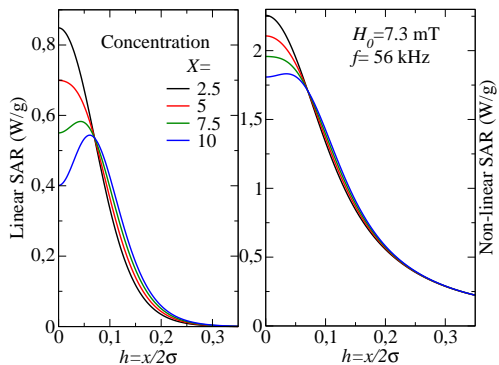}
\par\end{centering}
\caption{Same parameters as in Fig. \ref{fig:SAR_vs_dcField-hac=00003D3mT},
but for a higher AC field intensity $H_{0}=7.3{\rm \ mT}$.\label{fig:SAR_vs_dcField_hac=00003D7.3mT}}
\end{figure}

The two different scaling laws, ${\rm SAR^{\left(1\right)}}\propto H_{0}^{2}$
and ${\rm SAR^{\left(3\right)}}\propto H_{0}^{4}$, suggest that one
should be careful when using Eq. (\ref{eq:total_SAR}) to compute
the SAR for higher AC fields. In Fig. \ref{fig:SAR_vs_dcField_hac=00003D7.3mT}
we plot ${\rm SAR^{\left(1\right)}}$ and ${\rm SAR}={\rm SAR^{\left(1\right)}}+{\rm SAR^{\left(3\right)}}$
for $H_{0}=7.3{\rm \ mT}$. By comparing the left panels of Figs.
\ref{fig:SAR_vs_dcField-hac=00003D3mT} and \ref{fig:SAR_vs_dcField_hac=00003D7.3mT},
which display ${\rm SAR^{\left(1\right)}}$ for the two different
AC field amplitudes, 3 mT and 7.3 mT, we see that the curves are exactly
the same but merely rescaled by a global factor $\left(7.3/3\right)^{2}$.
This is in contrast with the respective right panels: upon including
the cubic correction in the susceptibility, the simple scaling no
longer applies. Indeed, a comparison of the left and right panels of Fig. \ref{fig:SAR_vs_dcField_hac=00003D7.3mT} shows that while the qualitative behavior remains
the same, the SAR is greatly enhanced by the correction ${\rm SAR^{\left(3\right)}}$.
This hints at the fact that it might be necessary to consider even
higher order terms {[}see Ref. \onlinecite{DejVerKach_springer20}{]}. 

\begin{figure}
\begin{centering}
\includegraphics[width=6.5cm]{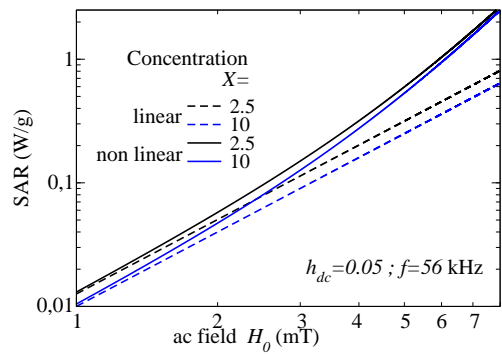}
\par\end{centering}
\caption{Log-Log plot of the linear and nonlinear SAR as a function of the
AC field intensity for two different concentrations $X=2.5$ and 10,
a fixed frequency and a finite DC field $h=0.05$.\label{fig:log_SAR_vs_acField}}
\end{figure}

On the other hand, this addresses another issue regarding the critical
product \citealp{brezovich88mpm} $H_{0}f\simeq4.85\times10^{8}\,{\rm A.m^{-1}.s^{-1}}$,
considered as a physiological limit. From the physicist standpoint,
this limit should be considered as an approximation of a more general
condition. Indeed, the expansion of Eq. (\ref{eq:total_SAR}) can
formally be rewritten as ${\rm SAR}\propto\left(H_{0}f\right)^{2}\left(1+\alpha H_{0}^{2}\right)$,
which thus shows that the product $H_{0}f$ is only relevant at very
low field intensities. Consequently, our results confirm the physiological
argument when the quadratic regime dominates the SAR. To illustrate
this, in Fig. \ref{fig:log_SAR_vs_acField} we plot the linear and
nonlinear SAR as a function of $H_{0}$ for two different concentrations.
In this log-log plot, the linear SAR exhibits, as expected, a linear
behavior, while the nonlinear SAR shows a crossover between a quadratic
(same slope as ${\rm SAR^{\left(1\right)}}$) and a quartic behavior
for higher fields. The slope rapidly changes and hardens as $H_{0}$
increases, with a correction that becomes very large near $H_{0}=7{\rm \ mT}$,
suggesting that higher order terms should become dominant. This
particular value of $H_{0}$ can be understood upon inspecting the
relative equilibrium susceptibilities at orders 1 and 3, since they
roughly give the orders of magnitudes of $\chi^{\left(1\right)}$
and $\chi^{\left(3\right)}$, as can be seen in Eqs. (\ref{eq:CubicSuscep})
and (\ref{eq:Debye}). The remaining terms of these equations are
related to the dynamics with the imaginary parts scaling as $\sim\eta$
in both cases. A rough estimate of the critical AC field value for
which the ${\rm SAR}$ goes to a quartic behavior, may
be inferred from the condition $H_{0}^{2}\left|\chi_{{\rm eq,free}}^{\left(3\right)}/\chi_{{\rm eq,free}}^{\left(1\right)}\right|\apprle1$.
This leads to $H_{0,c}\sim\frac{k_{B}T}{\mu_{0}m}$ which, for the
present sample evaluates to $7.2{\rm \ mT}$ and thereby $H_{0,c}f\simeq3.2\times10^{8}{\rm A.m^{-1}.s^{-1}}$. 

\begin{figure}
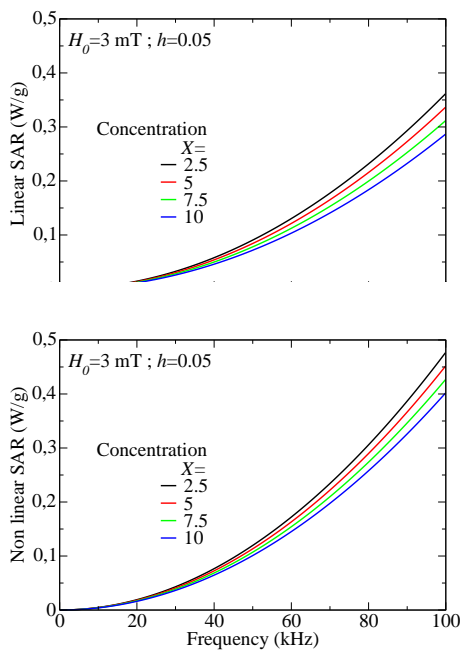

\begin{centering}
\includegraphics[width=6cm]{Fig5a.eps}
\par\end{centering}
\begin{centering}
\includegraphics[width=6cm]{Fig5b.eps}
\par\end{centering}
\caption{Linear and nonlinear SAR as a function of the AC field frequency for
two different concentrations, for a fixed AC field intensity $H_{0}=3{\rm \ mT}$
and a finite DC field $h=0.05$.\label{fig:SAR_vs_freq}}
\end{figure}

For fields below $H_{0,c}$, a criterion based on the quantity $\left(H_{0}f\right)^{2}$
makes sense both from a physiological and a physical point of view.
According to Eq. (\ref{eq:SAR_chi_ac}) the ${\rm SAR}$ is equal
to $H_{0}^{2}f$ multiplied by a contribution from the imaginary part
of the susceptibility $\chi^{\second}$. In terms of frequency, the
latter scales with $f$ at all orders of the expansion and, in terms
of the field intensity, the correction of order $\left(2n+1\right)$
scales with $H_{0}^{2n}$. Altogether, this implies that for fields
below $H_{0,c}$ the expansion is dominated by the first order, and
thus ${\rm SAR}\sim\mu_{0}\left(H_{0}f\right)^{2}$. This has been
checked by investigating the ${\rm SAR}$ as a function of the frequency
in Fig. \ref{fig:SAR_vs_freq}. For $H_{0}=3{\rm \ mT}$, the linear
and non linear SAR are qualitatively and quantitatively close and
they both exhibit a $f^{2}$ behavior since both $\Im\left[\chi^{\left(1\right)}\right]$
and $\Im\left[\chi^{\left(3\right)}\right]$ are proportional to the
frequency.

\section{Conclusion}

In the present investigation we have expressed the specific absorption
rate (SAR) in terms of the out-of-phase of the dynamical susceptibility
and have thereby provided analytical formulae that account for intrinsic
and collective effects, as well as the first nonlinear correction
to the AC susceptibility. In particular, the competition between the
DC field and dipolar interactions which, according to our previous
study, leads to a nonmonotonous behavior for the SAR, carries over,
at least qualitatively, to the nonlinear regime for the AC field intensity.
This implies that the DC field, even in the regime of higher AC field
intensities, remains a key parameter for optimizing the SAR for such
$2D$ arrays of nanomagnets.

Incidentally, we discussed applications to hyperthermia and the corresponding
(physiological) upper limit on the product $H_{0}f$. We have seen
that nonlinear contributions to the SAR bring an extra dependence on
the AC field intensity and frequency. As a consequence, it is more
relevant, from the physical point of view, and as long as the SAR
is used for assessing the efficiency of magnetic hyperthemia, to consider
instead the factor $\left(H_{0}f\right)^{2}$with which the SAR globally
scales.

\section*{Data Availability}
The data that support the findings of this study are available from the corresponding author upon reasonable request.
\newpage{}

\bibliography{Bibfile}

\end{document}